\documentclass[aps,prl,superscriptaddress,twocolumn,longbibliography,amssymb,amsmath]{revtex4-2}
\usepackage{graphicx}
\usepackage{dcolumn}
\usepackage{color}
\definecolor{darkblue}{rgb}{0,0.02,0.45}
\usepackage[colorlinks,bookmarks=false,citecolor=darkblue,linkcolor=red,urlcolor=blue]{hyperref}

\begin{document}
\title{Frustration-induced multiferroicity in hauerite MnS$_2$}

\author{Vilmos Kocsis}
\email{v.kocsis@ifw-dresden.de}
\affiliation{Leibniz Institute for Solid State and Materials Research Dresden, 01069 Dresden, Germany}

\author{Harish K. Singh}
\affiliation{Department of Chemistry and Physics of Materials, University of Salzburg, Jakob-Haringer-Strasse 2a, 5020 Salzburg, Austria}

\author{Kranthi K. Bestha}
\affiliation{Leibniz Institute for Solid State and Materials Research Dresden, 01069 Dresden, Germany}

\author{Yaqian Guo}
\affiliation{Leibniz Institute for Solid State and Materials Research Dresden, 01069 Dresden, Germany}

\author{Maxim Mostovoy}
\email{m.mostovoy@rug.nl}
\affiliation{Zernike Institute for Advanced Materials, University of Groningen, Nijenborgh 3, 9747 AG Groningen, The Netherlands}

\author{Jeroen van den Brink}
\email{j.van.den.brink@ifw-dresden.de}
\affiliation{Leibniz Institute for Solid State and Materials Research Dresden, 01069 Dresden, Germany}
\affiliation{W{\"u}rzburg-Dresden Cluster of Excellence ctd.qmat, 01069 Dresden, Germany}
\affiliation{Institute of Theoretical Physics, Technische Universit{\"a}t Dresden, 01062 Dresden, Germany}

\author{Oleg Janson}
\email{o.janson@ifw-dresden.de}
\affiliation{Leibniz Institute for Solid State and Materials Research Dresden, 01069 Dresden, Germany}

\begin{abstract}

Pyrite-type mineral hauerite MnS$_2$ is a magnetic insulator with localized
$S=\frac52$ moments. Below 48\,K, it develops a collinear antiferromagnetic
order, accompanied by unit-cell doubling and breaking the inversion symmetry.
We demonstrate that the latter gives rise to a ferroelectric polarization of up to 360\,$\mu$C/m$^2$ rendering MnS$_2$ a type-II multiferroic.
Microscopically, the electronic polarization originates from S$_2$ dimers that
acquire a dipole moment induced by the frustrated magnetic ordering of
neighboring Mn sites. We show that time-reversal symmetry leaves ferroelectric
domains intact, and that domain switching necessitates overcoming a high energy
barrier set by the magnetic exchange energy scale.  \end{abstract}

\maketitle

Multiferroic materials combine the properties of ferroelectrics and magnets and
are commonly divided into two groups~\cite{Khomskii06,Khomskii09}. In type-I
multiferroics, ferroelectricity (FE) and magnetism have different origins and
pertain to different active subsystems of a material. Their magnetic order
parameter, breaking the time-reversal symmetry $\mathcal{T}$, and ferroelectric
order parameter, breaking spatial inversion symmetry $\mathcal{P}$, coexist;
they may be coupled, but this coupling is typically weak.

Type-II multiferroics are different, because FE sets in at the magnetic
ordering transition and is underlain by $\mathcal{P}$-breaking in the ordered
phase. This may naturally happen if magnetic moments are twisted into a
spiral~\cite{Katsura05,Mostovoy06}. Another route for breaking inversion is
charge ordering~\cite{Brink08}, which occurs in complex oxides such as
rare-earth ($R$) manganites $R$Mn$_2$O$_5$ \cite{Hur04,Chapon06,Giovannetti09}.
In a more general context, certain combinations of nonsymmorphic symmetries
with the propagation vector naturally favor type-II
multiferroicity~\cite{perez-mato16}. Recently, Matsuda \textsl{et al.} studied
collinear antiferromagnets with a nonzero propagation vector as potential
multiferroics. In a direct calculation, they predicted that hauerite MnS$_2$
features nonzero Berry curvature and Berry curvature dipole, notably without
the spin-orbit coupling~\cite{matsuda25}.

\begin{figure}[tb]
\centering
\includegraphics[width=\columnwidth]{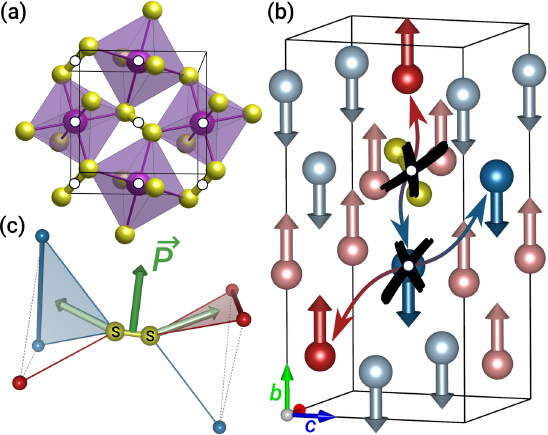}
\caption{Inversion symmetry breaking and emergence of dipole moments on S$_2$
dimers. (a) MnS$_6$ octahedra (violet) and S$_2$ dimers (yellow dumbbells) in
the pyrite-type cubic structure of paramagnetic MnS$_2$.  Inversion centers
(open circles) are located at Mn atoms and midpoints of S$_2$ dimers. (b) The
antiferromagnetic order doubles the unit cell and breaks $\mathcal{P}$ as
illustrated by curvilinear arrows connecting the loci of erstwhile inversion
centers (crossed-out open circles) with Mn atoms having antiparallel moments. 
(c) In the magnetically ordered state, both atoms of an S$_2$ dimer have three
Mn neighbors; two thereof have parallel moments, allowing for a dipole moment
on S (green arrow). The noncollinearity of electric dipole moments on two S atoms gives rise to
a dipole moment $\vec{P}$.}
\label{fig:str}
\end{figure}

In this Letter, we demonstrate that this simple binary material with a
pyrite-type crystal structure, discovered as a mineral in
1846~\cite{Anthony05}, exhibits ferroelectric polarization up to
360\,$\mu$C/m$^2$ in the antiferromagnetic state, and is thereby a type-II
multiferroic featuring neither charge ordering nor magnetic noncollinearity.
On the microscopic level, its multiferroicity is rooted in the frustrated
nature of the face-centered cubic (fcc) sublattice formed by the magnetic Mn atoms.
We show that in contrast to conventional antiferromagnets featuring a single
N{\'e}el vector, the magnetic ordering of MnS$_2$ is characterized by two
N{\'e}el vectors, describing two magnetic subsystems with frustrated interactions between them. Their interplay not only induces a ferroelectric polarization, but also adds a high energy barrier for domain switching, which is set by the Heisenberg
exchange scale.  This makes a key difference to conventional,
single-N{\'e}el-vector antiferromagnets, where the domain wall energy and width are
governed by relatively small magnetic anisotropies of relativistic origin.

\paragraph{Crystal and magnetic structure}
In the paramagnetic state, hauerite features a simple cubic pyrite-type
structure, with regular MnS$_6$ octahedra forming an fcc lattice interleaved by S$_2$ dumbbells
(Fig.~\ref{fig:str}, a).  Magnetic ordering doubles the cell in the $b$
direction~\cite{Corliss1958, Hastings1959}; the resulting magnetic structure
described by a cell with 8 Mn atoms and the propagation vector
$\vec{q}=(1,\!\frac12,\!0)$~\cite{Chattopadhyay1991PRB}.  The spin
model features dominant nearest-neighbor exchanges between Mn $S=\frac52$
moments, and an order of magnitude weaker interactions between second
neighbors~\cite{Chatterji2019}; the interplay between these frustrated
interactions stabilizes a collinear ground state~\cite{Schick2020}.
Importantly, this magnetic ordering of MnS$_2$ breaks $\mathcal{P}$ of the
pyrite structure (Fig.~\ref{fig:str}, b), allowing the S$_2$ dimers -- as we
will show below -- to acquire dipole moments (Fig.~\ref{fig:str}, c) whose
mutual arrangement gives rise to a net ferroelectric polarization. 

\begin{figure}[tb]
\centering
\includegraphics[width=0.49\textwidth]{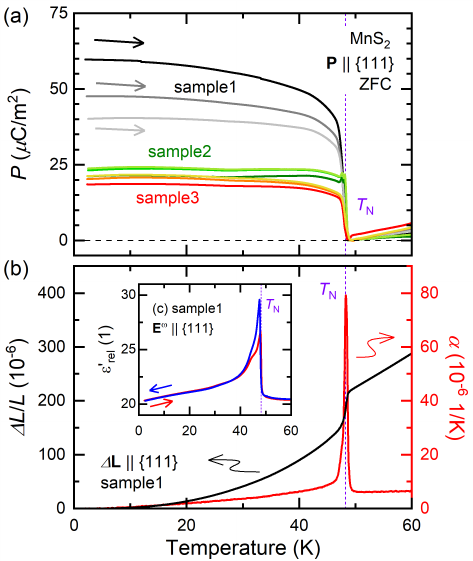}
\caption{Temperature dependence of the (a) zero-field-cooled ferroelectric
polarization ($P$) and (b)
the relative length change ($\Delta{L}/L$) and the thermal expansion
coefficient ($\alpha$). Inset (c): Real part of the dielectric constant ($\epsilon’$) 
of MnS$_2$.}
\label{fig:exp}
\end{figure}

\begin{figure}[tb]
\centering
\includegraphics[width=0.49\textwidth]{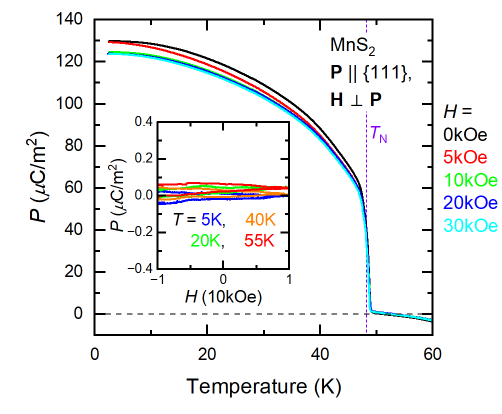}
\caption{Temperature dependence of the ferroelectric polarization $P$ along
$[111]$ at selected fields.  Note that polarization in MnS$_2$ is confined to the $c$ axis, and hence the intrinsic value of $P$ is $\sqrt{3}$ times larger.
Inset: Magnetic field dependence of the polarization at selected temperatures.}
\label{fig:p_field}
\end{figure}

\paragraph{Ferroelectric polarization and dielectric constant measurements}
We used a naturally grown single crystal of hauerite with triangular-shaped
$\{111\}$ faces taken from the Geoscientific Collections of the TU Bergakademie
Freiberg, sample number LaSa40300.  Ferroelectric polarization was measured
using the charge collection mode of an electrometer (6517A, Keithley) in a
Physical Property Measurement System (PPMS, Quantum Design).  The samples were
thoroughly cleaned and polished, and measurement electrodes were painted
2.8\,mm apart on the opposing $\{111\}$ normal faces using silver paint with an
$\sim$2.9\,mm$^2$ cross-section. 
First, three samples were cooled to $T$ = 2\,K in the absence of any poling electric field (ZFC) to prepare a multi-domain measurement without the possibility of trapped charges.
Next, we measured polarization in the warming runs, as shown in Fig.~\ref{fig:exp}(a).
The observed polarization showed a $\Delta{P}$\,$\approx$\,25-50\,$\mu${C}/m$^2$ jump at $T_{\text{N}}$, indicating the emergence of ferroelectricity. Importantly, these measurements give no information on the saturation value of the polarization $P_{\text{sat}}$: for instance, with a very low poling electric and magnetic field ($E_0$ = 0.107\,kV/cm), we were able to reach $\sim$120 \,$\mu$C/m$^2$ (Fig.~\ref{fig:p_field}). Since these experiments can be affected by trapped charges, we performed additional
field-cooled measurements~\cite{suppl}
that yielded the saturation value along the [111] axis is $P_{\text{sat}}^{[111]}$\,$\approx$\,210\,$\mu$C/m$^2$, which is a projection of the actual polarization constrained to the $c$ axis of the crystal. Therefore, we can estimate the saturation value of the polarization as $P_{\text{sat}}$\,$\approx$\,360\,$\mu$C/m$^2$.

Measurements in a magnetic field revealed that the polarization has no magnetic field dependence. This behavior is expected
for the magnetic space group $P_bca2_1$ (29.105), whose symmetry forbids linear
and quadratic magnetoelectric effects. To further demonstrate ferroelectricity at
the magnetic ordering, we measured the dielectric constant using a capacitance
bridge (2550A, Andeen-Hagerling).  At $T_{\text{N}}$, the real part
($\varepsilon'$) shows a sharp peak with 150$\%$ magnitude and a broad
hysteresis indicative of a first-order transition~\cite{Chattopadhyay1991PRB}.

\paragraph{Thermal expansion} The hysteretic behavior of $\epsilon'$ may hint
at a structural distortion. Indeed, a three-dimensional spin-Peierls transition
scenario has been discussed for MnS$_2$ based on high-quality synchrotron
diffraction data~\cite{Kimber2015}, yet the magnitude of the reported
pseudo-tetragonal distortion ($c$/$a$ = 1.0006) makes detection of this
distortion very challenging. To get more insight into the transition, we
measured thermal expansion in the $[111]$ direction on warming at 0.25\,K/min
using a high-resolution capacitance dilatometer (mini, Kuechler Innovative
Measurement Technology), integrated into the PPMS and the capacitance bridge
(2700A, Andeen-Hagerling), using a 2.5\,mm thick platelet as a reference.  At
$T_{\rm N}$, the thermal expansion coefficient develops a sizable peak
(Fig.~\ref{fig:exp}, b), indicating the magnetoelastic origin of the
ferroelectric polarization.  Using the Clausius-Clapeyron equation, we can
access the uniaxial-pressure dependence of $\partial{}T_{\rm
N}/\partial{}p_{[111]}$ = $V_{mol}\frac{\Delta{(\Delta{L}_{111}/{L}_{111})}}{\Delta{S}_{magn}}$ = 0.7(1)\,K/GPa, which is approximately six times
smaller than that of $\partial{}T_{\rm N}/\partial{}p$ = 4.8\,K/GPa concluded
from hydrostatic-pressure studies~\cite{Westrum1970,Sidorov2018JETP,suppl}. Based on this
difference, we conclude that the magnetic ordering transition in MnS$_2$ is
accompanied by a small, yet significantly anisotropic contraction.

\paragraph{First-principles calculations} 
To understand the mechanism behind the multiferroicity of MnS$_2$, we turn to
density-functional-theory (DFT) calculations. (For readability, we defer all
computational details to the End Matter section.) Previous high-precision
synchrotron measurements revealed a splitting of the $(511)$ reflection, yet
attempts to refine the structure in the orthorhombic $Pbca$ space group yielded
only a minute distortion~\cite{Kimber2015}. To clarify whether a structural
distortion takes place, we performed a full relaxation of the MnS$_2$ structure
in its magnetically ordered state.  In this way, we find that a distorted
structure described by the orthorhombic space group $Pca2_1$ has 4.8\,meV/f.u.\
lower total energy than the undistorted phase; the maximal atomic displacement
amounts to $3.6\times10^{-3}$\,\r{A}. Such displacements can be too small to be
reliably detected with standard diffraction methods. To overcome this
limitation, we suggest resorting to resonant elastic X-ray scattering (REXS), an
element-specific technique, which is particularly suited for studying weak
symmetry breaking. However, the presence of such distortions is further
underpinned by calculations of phonon spectra: the polar $Pca2_1$ structure is
dynamically stable, while the undistorted phase -- metrically equivalent to the
cubic $Pa\bar{3}$ phase -- shows imaginary modes.  This agrees with our thermal
expansion measurements showing a contraction at $T_{\text{N}}$, and further
indicates that the antiferromagnetism of MnS$_2$ should be addressed
computationally using  the $Pca2_1$ structural input rather than the
undistorted -- metrically cubic -- structure.

\begin{figure}[t]
\centering
\includegraphics[width=\columnwidth]{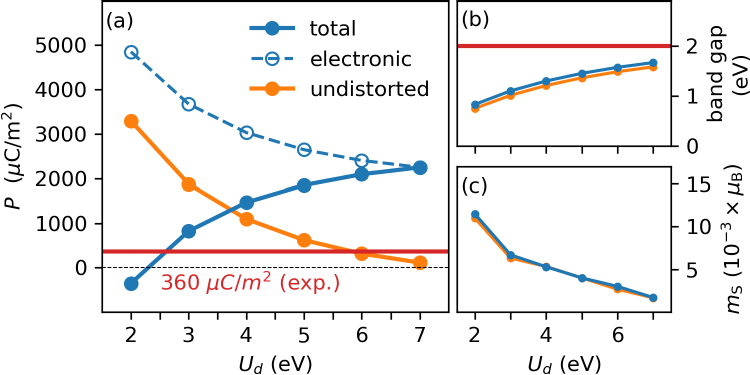}
\caption{Results of GGA+$U$ calculations. (a) Total ferroelectric polarization
$P$ (blue solid) and its electronic contribution (blue dashed) as a function of
the Coulomb repulsion parameter $U_d$. For reference, we plot the polarization
of the undistorted structure (orange) which is fully of electronic origin. We
further plot the $U_d$ dependence of the band gap (b) and the average absolute magnetic
moment on S atoms (c). Experimental estimates are indicated with red lines.}
\label{fig:U-dependence}
\end{figure}

We are now in position to calculate the electric polarization in the
magnetically ordered phase within DFT+$U$. To reproduce the insulating state of
MnS$_2$, we use the DFT+$U$ method and fix the Hund exchange $J_d$ in the Mn
$3d$ shell to 1\,eV, a common value for $3d$ systems. For the Coulomb repulsion
parameter $U_d$, previous DFT+$U$ studies of MnS$_2$ employed $U_d$ =
3\,eV~\cite{Persson2006, Kimber2014} which yields a band gap of about 1\,eV,
thereby strongly underestimating the experimental value of
$\sim2$\,eV~\cite{Zhou2024}.  To ensure our results are not artifacts of the
specific choice of $U_d$, we systematically varied $U_d$ in a broad range from
2 to 7 eV~\footnote{We note that even for $U_d$ = 7\,eV (unusually large for a
middle-transition-metal element like Mn), the experimental band gap is still
underestimated by about 15\% (Fig.~\ref{fig:U-dependence}, b). While this
discrepancy deserves a further investigation, it does not affect our central conclusions.}.

The results of our DFT+$U$ calculations as a function of $U_d$ are shown in
Fig.~\ref{fig:U-dependence}. For low $U_d$ values, the total polarization is
small, because its ionic ($P^{\text{ion}}$) and electronic ($P^{\text{el}}$)
contributions are well balanced and nearly cancel each other: both amount to
$\sim$4500--5000\,$\mu$C/m$^2$ with opposite sign.  For larger $U_d$ values the
electronic polarization (dashed curve in Fig.~\ref{fig:U-dependence}, a) is
drastically reduced, and the total polarization (solid curve in
Fig.~\ref{fig:U-dependence}, a) jumps to values exceeding $\sim$2000
$\mu$C/m$^2$. For comparison, we calculated the ferroelectric polarization of
the undistorted (metrically cubic) structure, where the ionic contribution is
strictly absent; the resulting $P(U_d)$ dependence (orange curve in
Fig.~\ref{fig:U-dependence}, a) closely resembles $P^{\text{el}}$ of the
distorted structure, albeit with a different offset. This similarity leads us
to the conclusion that the microscopic mechanism behind $P^{\text{el}}$ can not
be driven by structural distortions.

\paragraph{Microscopic model of multiferroic MnS$_2$} To identify the root
cause of multiferroicity, we focus on hybridization between Mn $d$ and S $p$
states of neighboring atoms, which is known to be crucial for electric
polarization and strongly affected by $U_d$~\cite{Giovannetti08}. The corollary
of such hybridization is a partial transfer of magnetization from a magnetic
atom (Mn) to a ligand (S). In many antiferromagnets this transfer is fully
compensated by the contribution of the other Mn neighbor, resulting in zero net
magnetization of the ligand atom.  This is very different in MnS$_2$, where
each S ligand is shared by three Mn neighbors. Moreover, in the magnetically
ordered state each S$_2$ dumbbell has a distinct structure: one of its S atoms has
two spin-up and one spin-down Mn neighbors, while the other has one spin-up and
two spin-down (Fig.~\ref{fig:str}, c). As a result, each individual S atom
inherits a small, yet finite magnetization from the two neighboring Mn atoms
with parallel magnetic moments. This magnetization is fully compensated by the other S atom within the dumbbell.

\begin{figure}[tb]
\centering
\includegraphics[width=8.6cm]{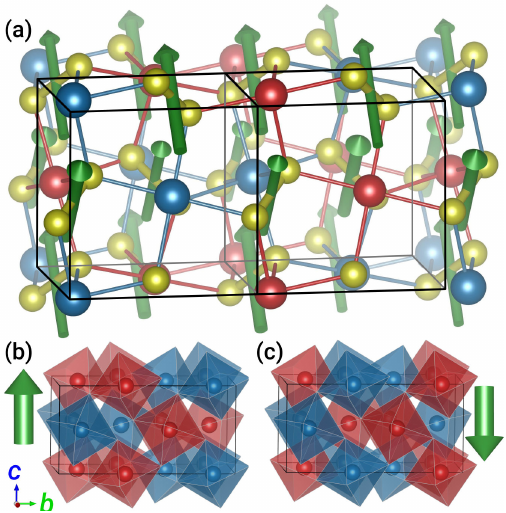}
\caption{
Ferroelectricity of MnS$_2$. (a) Spin-up (red) and spin-down (blue) Mn atoms in
the magnetic structure of MnS$_2$.  Polarization of each S$_2$ dimer is
indicated by a green arrow; net electric polarization is along $c$. (b) and (c)
Two magnetic domain structures with opposite polarization.  Only magnetic
MnS$_6$ octahedra are shown.  Slicing both domains in the $ac$ plane (each
panel depicts four such slices) reveals that each second slice has opposite
orientation in the two magnetic domains.}
\label{fig:domains}
\end{figure}

If we now inspect the average absolute magnetization of S atoms, it shows a
strong $U$ dependence (Fig.~\ref{fig:U-dependence}, c), largely resembling that
of $P^{\text{el}}$.  This similarity allows us to rationalize the
multiferroicity of MnS$_2$ within a simple, yet microscopic model, whose basic
element is the S$_2$ dumbbell. Since the $d-p$ hybridization will be different
for same- and opposite-spin Mn-S pairs, we can approximately assume that each S
atom features a dipole moment aligned towards the respective Mn-Mn edge
(Fig.~\ref{fig:str}, c). By plotting two dipoles of the same S$_2$
dumbbell, we immediately see that the two vectors are not antiparallel, and
hence each dumbbell has a finite polarization. Most importantly, the summation
over all 8 dumbbells within the magnetic unit cell gives a nonzero overall
polarization along $c$ (Fig.~\ref{fig:domains}, a) -- the only symmetry-allowed
direction of electronic polarization.

\paragraph{Magnetoelectric coupling} The relation between multiferroicity  and magnetic frustration in MnS$_2$ can be understood from a symmetry analysis of the magnetoelectric coupling (see End Matter for details). We introduce two
N\'eel vectors, $\mathbf{L}_1$ and $\mathbf{L}_2$, describing the antiferromagnetic spin ordering in odd and even $ac$ planes with the wave vector $(0,\frac{1}{2},0)$ in the $Pa\bar{3}$ setting. With $\mathbf{L}_1$ being odd and $\mathbf{L}_2$ being even under inversion, any linear coupling between the two magnetic subsystems is forbidden -- a sign of frustration. On the other hand, the magnetoelectric coupling
$P_a \left(\mathbf{L}_1\cdot\mathbf{L}_2\right)$ is allowed by crystal and time-reversal symmetries ($P\|c$ in the $P_bca2_1$ setting). The polarization sign depends on whether $\mathbf{L}_1$ and $\mathbf{L}_2$ are parallel or antiparallel, while the magnetic energy is independent of it. Thus, ferroelectricity spontaneously lifts magnetic frustration. Invariance of the scalar product $\left(\mathbf{L}_1\cdot\mathbf{L}_2\right)$ under arbitrary rotations in spin space indicates a non-relativistic origin of the electric polarization, consistent with the large value of $P$ found in our calculations.

\paragraph{Ferroelectric domains} To construct a ferroelectric domain with opposite polarization, we need to flip
half of the magnetic moments: in panels (b) and (c) of Fig.~\ref{fig:domains}
these are moments in every second slice in the $ac$ plane. This result is
readily confirmed by DFT+$U$ calculations that yield polarizations of equal
magnitude, but opposite sign. Thus, domain reorientation incurs a substantial
energy penalty governed by the leading magnetic exchange coupling.  To estimate
this energy, we take one ferroelectric domain (Fig.~\ref{fig:domains}, b) and
rotate magnetic moments in each second $ac$ plane. We consider
different rotation angles up to 180$^{\circ}$, for which the other ferroelectric
domain (Fig.~\ref{fig:domains}, c) is recovered.  In this way, we find that the
energy barrier for switching between different ferroelectric domains amounts to
$\sim$1\,meV per formula unit, which is sufficiently high to prevent domain
flipping at low temperatures. Crucially, this energy separation is driven by
the Heisenberg exchange and not by magnetic anisotropy. This makes the  ferroelectric domain wall, in which $\mathbf{L}_1$ rotates with respect to $\mathbf{L}_2$ through $180^\circ$, very sharp, as in proper ferroelectrics.

\paragraph{Summary}
We demonstrated that the collinear antiferromagnet hauerite MnS$_2$ exhibits
type-II multiferroicity below its magnetic ordering transition. In electric
polarization measurements, we found a ferroelectric polarization reaching up to
360 $\mu\text{C/m}^2$, accompanied by anomalies in the dielectric constant and
thermal expansion at $T_{\text{N}}$.  In contrast to multiferroics driven by
spin-spirals or charge-ordering, the polarization in MnS$_2$ has an electronic
origin, arising from local dipole moments on the S$_2$ dimers through
spin-dependent $d$--$p$ hybridization.  By performing a symmetry analysis and
density-functional-theory calculations, we confirmed that ferroelectric domains
remain intact under time reversal, and estimated the switching energy barrier
to reach 1 meV per formula unit.  These findings establish collinear,
frustrated magnets as a highly promising route toward multiferroic systems with
domain flipping protected by the energy scale set by the isotropic magnetic
exchange.

\begin{acknowledgments}
\paragraph{Acknowledgments} The authors acknowledge fruitful discussions with
Sang-Wook Cheong, Anja U.\ B.\ Wolter-Giraud, and Christin Kehrer. VK was
supported by the Alexander von Humboldt Foundation. The authors are grateful to
Gerhard Heide from the Geoscientific Collections of the TU Bergakademie
Freiberg for providing hauerite samples. HKS acknowledges the EuroHPC Joint
Undertaking for granting access to the VEGA supercomputer at IZUM, Slovenia,
through project EHPC-REG-2025R01-197-EuroHPC.  HKS thanks Ulrich Aschauer for the discussion. YG and OJ thank Ulrike Nitzsche
for technical assistance. We acknowledge financial support from the German
Forschungsgemeinschaft (DFG, German Research Foundation) through Individual Research Grant (project-id 540912241), W{\"u}rzburg-Dresden Cluster of Excellence on Complexity, Topology and Dynamics
in Quantum Matter – ctd.qmat (EXC 2147, project-id 390858490) and through SFB
1143 (project-id 247310070).
\end{acknowledgments}

\newpage
\appendix
\section*{End Matter}
\subsection*{Details of DFT calculations}
\label{sec:dft_method}

For DFT calculations, we used the projector-augmented-wave code VASP version
6.4.2~\cite{vasp_1, vasp_2} with standard pseudopotentials~\cite{vasp_paw}. 
An energy cutoff of 600 eV was applied to the plane-wave basis, and the
Brillouin zone was sampled using a Monkhorst-Pack $8\times4\times8$ mesh of
$k$-points.  The Gaussian method was used to determine the partial occupancies
of orbitals with a smearing width of 0.01 eV. The exchange-correlation
functional was parameterized by the revised Perdew-Burke-Ernzerhof (PBEsol)
form within the generalized gradient approximation (GGA) plus U method
\cite{perdew2008restoring}. 

We treated the localized Mn \(d\) states using the Dudarev
approach~\cite{dudarev1998electron}, varying the on-site Coulomb parameter
\(U\) from \(0.0\) to \(8.0\,\mathrm{eV}\). Phonon dispersions were evaluated
in the harmonic approximation using the finite-displacement (frozen-phonon)
method with a \(2\times2\times2\) supercell and analyzed with
Phonopy~\cite{phonopy_1, phonopy_2}.  The magnetic space group was calculated using
the FINDSYM software \cite{stokes2005findsym}.

To calculate the intrinsic ferroelectric polarization, we used the Berry-phase
method~\cite{resta1994macroscopic} implemented in VASP.

\subsection*{Symmetry analysis}
\label{sec:symmetry_analysis}

A phenomenological analysis of the magnetoelectric coupling in MnS$_2$ is done using magnetic order parameters and crystal symmetries in the paramagnetic state (space group $Pa\bar{3}$). The magnetic Mn$^{2+}$ ions in the unit cell, labeled from 1 to 4, occupy 4a positions with the coordinates: $\mathbf{r}_1 = (0,1/2,1/2)$, $\mathbf{r}_2 = (1/2,0,1/2)$, $\mathbf{r}_3 = (1/2,1/2,0)$ and $\mathbf{r}_4 = (0,0,0)$. Their transformations under generators of $Pa\bar{3}$ are shown in Table~\ref{tab:MnS2sites}.

\begin{table}[htbp]
        \centering
        \begin{tabular}{|c|c|c|c|c|c|}
                \hline
                Site & Coordinates    &$3_{[111]}$&$\tilde{2}_{[010]}$&$\tilde{2}_{[001]}$& $I$\\
                [0.4ex]
                \hline
                $1$ & $(0, 1/2, 1/2)$ &$2$&$4+b$&$3-b+c$&$1-b-c$\\
                [0.4ex]
                $2$ & $(1/2,0, 1/2)$  &$3$&$3-a$&$4+c$  &$2-a-c$\\
                [0.4ex]
                $3$ & $(1/2,1/2,0)$   &$1$&$2-a+b$&$1-b$  &$3-a-b$\\
                [0.4ex]
                $4$ & $(0, 0, 0)$     &$4$&$1$&$2$    &$4$\\
                [0.4ex]
                \hline
        \end{tabular}
        \caption{Transformations of the $4a$ sites under the generators of $Pa\bar{3}$ space group: $3_{[111]} = (z,x,y), \tilde{2}_{[010]} =  (-x,y+1/2,-z+1/2), \tilde{2}_{[001]} = (1/2-x,-y,z+1/2)$ and $I = (-x,-y,-z)$.
                Site $2-a+b$ is located at $\mathbf{r}_2 - \mathbf{a} + \mathbf{b}$.}
        \label{tab:MnS2sites}
\end{table}

The magnetically ordered state with the wave vector $\mathbf{Q} = (0,1/2,0)$ is described by the two-component N\'eel vector:
\begin{equation}
\mathbf{L} = 
\left(\begin{array}{c}
\mathbf{L}_1\\
\mathbf{L}_2
\end{array}\right)
=
\left(\begin{array}{c}
\mathbf{S}_{1l}-\mathbf{S}_{3l}\\
\mathbf{S}_{2l}-\mathbf{S}_{4l}
\end{array}\right) e^{2 \pi i \mathbf{Q}\cdot\mathbf{R}_l},
\label{eq:MnS2Neel}
\end{equation}
where $\mathbf{S}_{nl}$ is the spin at site $n=1,2,3,4$ of the unit cell $l$, and $\mathbf{R}_l$ is the coordinate of the unit cell. $\mathbf{L}_1$ and $\mathbf{L}_2$ describe AFM orders in neighboring $ac$-planes. Their transformation properties are shown in Table~\ref{tab:MnS2Neel}. The $120^{\circ}$-rotation around body diagonals is not considered, since it changes the direction of  $\mathbf{Q}$. Note that for the choice of $\mathbf{Q}$ used in this analysis, symmetry requires $P \parallel a$; by a cyclic permutation of axes this state can be transformed into the standard setting of space group $Pca2_1$ with $P \parallel c$ used in the main text.
\begin{table}[htbp]
        \centering
        \begin{tabular}{|c|c|c|c|}
                \hline
                &$\tilde{2}_{[010]}$& $\tilde{2}_{[001]}$& $I$\\
                [0.4ex]
                \hline
                $\left(\begin{array}{c}
                \mathbf{L}_1\\
                \mathbf{L}_2
                \end{array}\right)$& $R_y(\pi)
                \left(\begin{array}{c}
                -\mathbf{L}_2\\
                +\mathbf{L}_1
                \end{array}\right)$& $R_z(\pi) 
                \left(\begin{array}{c}
                +\mathbf{L}_1\\
                -\mathbf{L}_2
                \end{array}\right)$&
                $\left(\begin{array}{c}
                -\mathbf{L}_1\\
                +\mathbf{L}_2
                \end{array}\right)$\\
                [0.4ex]
                \hline
        \end{tabular}
        \caption{Transformations of the two-component N\'eel vector $\mathbf{L}$ under symmetry operations of $Pa\bar{3}$ space group. The $3\times3$ matrices $R_y(\pi)$ and $R_z(\pi)$ describe $180^\circ$-rotations around the $y$ and $z$ axes, respectively.
        }
        \label{tab:MnS2Neel}
\end{table}
Since $\mathbf{L}_1$ is odd under inversion and $\mathbf{L}_2$ is even,  the term  $\left(\mathbf{L}_{1}\cdot\mathbf{L}_{2} \right)$ in free energy is forbidden, showing that interactions of the two magnetic sublattices are frustrated. On the other hand, the magnetoelectric coupling,
\begin{equation}
H_{\rm me } = - \lambda E_x 
\left(
\mathbf{L}_{1}\cdot\mathbf{L}_{2}
\right),
\label{eq:me4}
\end{equation}
is symmetry allowed.

A more detailed understanding of the magnetoelectric coupling is achieved by considering electric dipole moments on Mn-Mn bonds that cancel each other in the paramagnetic state, but become uncompensated below N\'eel temperature. Here, we only consider nearest-neighbor bonds.
Using relations between the electric dipoles on symmetry-related bonds, we obtain three independent magnetoelectric coupling terms of non-relativistic origin invariant under all symmetry operations:
\begin{align}
&H_{\rm me}^{(1)} = \nonumber \\
&-g_1 E_x\left[\right.(2,4)-(2,4+a+c)+(4+a,2)-(4+c,2)\nonumber\\
&+(1+a,3+c)-(1,3)+(3+c,1)-(3,1+a)\left.\right]\nonumber\\
&-g_1 E_y\left[\right.(3,4)-(3,4+a+b)+(4+b,3)-(4+a,3)\nonumber\\
&+(2+b,1+a)-(2,1)+(1+a,2)-(1,2+b)\left.\right] \nonumber\\
&-g_1 E_z\left[\right.(1,4)-(1,4+b+c)+(4+c,1)-(4+b,1)\nonumber\\
&+(3+c,2+b)-(3,2)+(2+b,3)-(2,3+c)\left.\right],
\end{align}
\begin{align}
&H_{\rm me}^{(2)} = \nonumber \\
&-g_2 E_x\left[\right.(2,3)-(2+b,3+c)+(3+c,2)-(3,2+b) \nonumber\\
&+(1,4+c)-(1,4+b)+(4,1)-(4+b+c,1)\left.\right] \nonumber\\
&-g_2 E_y\left[\right.(3,1)-(3+c,1+a)+(1+a,3)-(1,3+c) \nonumber\\
&+(2,4+a)-(2,4+c)+(4,2)-(4+a+c,1)\left.\right] \nonumber\\
&-g_2 E_z\left[\right.(1,2)-(1+a,2+b)+(2+b,1)-(2,1+a) \nonumber\\
&+(3,4+b)-(3,4+a)+(4,3)-(4+a+b,2)\left.\right],
\end{align}
\begin{align}
&H_{\rm me}^{(3)} = \nonumber \\
&-g_3 E_x\left[\right.(1,2)-(1+a,2+b)+(2,1+a)-(2+b,1) \nonumber\\
&+(3,4+a)-(3,4+b)+(4,3)-(4+a+b,3)\left.\right] \nonumber\\
&-g_3 E_y\left[\right.(2,3)-(2+b,3+c)+(3,2+b)-(3+c,2) \nonumber\\
&+(1,4+b)-(1,4+c)+(4,1)-(4+b+c,1)\left.\right] \nonumber\\
&-g_3 E_z\left[\right.(3,1)-(3+c,1+a)+(1,3+c)-(1+a,3) \nonumber\\
&+(2,4+c)-(2,4+a)+(4,2)-(4+a+c,2)\left.\right].
\end{align}
The terms in parentheses are scalar products of spins, e.g. $(1+a,2+b) = \left(
\mathbf{S}_{1,l+a}\cdot\mathbf{S}_{2,l+b}
\right)$, where $\mathbf{S}_{1,l+a}$ denotes the spin of ion 1 in the unit cell $(l_x+1,l_y,l_z)$.

For the antiferromagnetic ordering with the wave vector $\mathbf{Q} = (0,1/2,0)$ and  $\left(
\mathbf{S}_{1}\cdot\mathbf{S}_{2} \right) = \left(
\mathbf{S}_{3}\cdot\mathbf{S}_{4} \right) = - \left(
\mathbf{S}_{1}\cdot\mathbf{S}_{3} \right) = - \left(
\mathbf{S}_{2}\cdot\mathbf{S}_{4} \right) = - \left(
\mathbf{S}_{1}\cdot\mathbf{S}_{4} \right) = - \left(
\mathbf{S}_{2}\cdot\mathbf{S}_{3} \right)$ \cite{Hastings1959}, the second and third magnetoelectric coupling terms give rise to a nonzero electric polarization along the $a$ axis ($P_b = P_c = 0$), in agreement with Eq.(\ref{eq:me4}).


\begin{thebibliography}{34}%
\makeatletter
\providecommand \@ifxundefined [1]{%
 \@ifx{#1\undefined}
}%
\providecommand \@ifnum [1]{%
 \ifnum #1\expandafter \@firstoftwo
 \else \expandafter \@secondoftwo
 \fi
}%
\providecommand \@ifx [1]{%
 \ifx #1\expandafter \@firstoftwo
 \else \expandafter \@secondoftwo
 \fi
}%
\providecommand \natexlab [1]{#1}%
\providecommand \enquote  [1]{``#1''}%
\providecommand \bibnamefont  [1]{#1}%
\providecommand \bibfnamefont [1]{#1}%
\providecommand \citenamefont [1]{#1}%
\providecommand \href@noop [0]{\@secondoftwo}%
\providecommand \href [0]{\begingroup \@sanitize@url \@href}%
\providecommand \@href[1]{\@@startlink{#1}\@@href}%
\providecommand \@@href[1]{\endgroup#1\@@endlink}%
\providecommand \@sanitize@url [0]{\catcode `\\12\catcode `\$12\catcode
  `\&12\catcode `\#12\catcode `\^12\catcode `\_12\catcode `\%12\relax}%
\providecommand \@@startlink[1]{}%
\providecommand \@@endlink[0]{}%
\providecommand \url  [0]{\begingroup\@sanitize@url \@url }%
\providecommand \@url [1]{\endgroup\@href {#1}{\urlprefix }}%
\providecommand \urlprefix  [0]{URL }%
\providecommand \Eprint [0]{\href }%
\providecommand \doibase [0]{https://doi.org/}%
\providecommand \selectlanguage [0]{\@gobble}%
\providecommand \bibinfo  [0]{\@secondoftwo}%
\providecommand \bibfield  [0]{\@secondoftwo}%
\providecommand \translation [1]{[#1]}%
\providecommand \BibitemOpen [0]{}%
\providecommand \bibitemStop [0]{}%
\providecommand \bibitemNoStop [0]{.\EOS\space}%
\providecommand \EOS [0]{\spacefactor3000\relax}%
\providecommand \BibitemShut  [1]{\csname bibitem#1\endcsname}%
\let\auto@bib@innerbib\@empty
\bibitem [{\citenamefont {Khomskii}(2006)}]{Khomskii06}%
  \BibitemOpen
  \bibfield  {author} {\bibinfo {author} {\bibfnamefont {D.}~\bibnamefont
  {Khomskii}},\ }\href
  {https://doi.org/https://doi.org/10.1016/j.jmmm.2006.01.238} {\bibfield
  {journal} {\bibinfo  {journal} {J. Magn. Magn. Mater.}\ }\textbf {\bibinfo
  {volume} {306}},\ \bibinfo {pages} {1} (\bibinfo {year} {2006})}\BibitemShut
  {NoStop}%
\bibitem [{\citenamefont {Khomskii}(2009)}]{Khomskii09}%
  \BibitemOpen
  \bibfield  {author} {\bibinfo {author} {\bibfnamefont {D.}~\bibnamefont
  {Khomskii}},\ }\href {https://doi.org/10.1103/Physics.2.20} {\bibfield
  {journal} {\bibinfo  {journal} {Physics}\ }\textbf {\bibinfo {volume} {2}},\
  \bibinfo {pages} {20} (\bibinfo {year} {2009})}\BibitemShut {NoStop}%
\bibitem [{\citenamefont {Katsura}\ \emph {et~al.}(2005)\citenamefont
  {Katsura}, \citenamefont {Nagaosa},\ and\ \citenamefont
  {Balatsky}}]{Katsura05}%
  \BibitemOpen
  \bibfield  {author} {\bibinfo {author} {\bibfnamefont {H.}~\bibnamefont
  {Katsura}}, \bibinfo {author} {\bibfnamefont {N.}~\bibnamefont {Nagaosa}},\
  and\ \bibinfo {author} {\bibfnamefont {A.~V.}\ \bibnamefont {Balatsky}},\
  }\href {https://doi.org/10.1103/PhysRevLett.95.057205} {\bibfield  {journal}
  {\bibinfo  {journal} {Phys. Rev. Lett.}\ }\textbf {\bibinfo {volume} {95}},\
  \bibinfo {pages} {057205} (\bibinfo {year} {2005})}\BibitemShut {NoStop}%
\bibitem [{\citenamefont {Mostovoy}(2006)}]{Mostovoy06}%
  \BibitemOpen
  \bibfield  {author} {\bibinfo {author} {\bibfnamefont {M.}~\bibnamefont
  {Mostovoy}},\ }\href {https://doi.org/10.1103/PhysRevLett.96.067601}
  {\bibfield  {journal} {\bibinfo  {journal} {Phys. Rev. Lett.}\ }\textbf
  {\bibinfo {volume} {96}},\ \bibinfo {pages} {067601} (\bibinfo {year}
  {2006})}\BibitemShut {NoStop}%
\bibitem [{\citenamefont {van~den Brink}\ and\ \citenamefont
  {Khomskii}(2008)}]{Brink08}%
  \BibitemOpen
  \bibfield  {author} {\bibinfo {author} {\bibfnamefont {J.}~\bibnamefont
  {van~den Brink}}\ and\ \bibinfo {author} {\bibfnamefont {D.~I.}\ \bibnamefont
  {Khomskii}},\ }\href {https://doi.org/10.1088/0953-8984/20/43/434217}
  {\bibfield  {journal} {\bibinfo  {journal} {J. Phys.: Condens. Matter}\
  }\textbf {\bibinfo {volume} {20}},\ \bibinfo {pages} {434217} (\bibinfo
  {year} {2008})}\BibitemShut {NoStop}%
\bibitem [{\citenamefont {{Hur}}\ \emph {et~al.}(2004)\citenamefont {{Hur}},
  \citenamefont {{Park}}, \citenamefont {{Sharma}}, \citenamefont {{Ahn}},
  \citenamefont {{Guha}},\ and\ \citenamefont {{Cheong}}}]{Hur04}%
  \BibitemOpen
  \bibfield  {author} {\bibinfo {author} {\bibfnamefont {N.}~\bibnamefont
  {{Hur}}}, \bibinfo {author} {\bibfnamefont {S.}~\bibnamefont {{Park}}},
  \bibinfo {author} {\bibfnamefont {P.~A.}\ \bibnamefont {{Sharma}}}, \bibinfo
  {author} {\bibfnamefont {J.~S.}\ \bibnamefont {{Ahn}}}, \bibinfo {author}
  {\bibfnamefont {S.}~\bibnamefont {{Guha}}},\ and\ \bibinfo {author}
  {\bibfnamefont {S.~W.}\ \bibnamefont {{Cheong}}},\ }\href
  {https://doi.org/10.1038/nature02572} {\bibfield  {journal} {\bibinfo
  {journal} {\nat}\ }\textbf {\bibinfo {volume} {429}},\ \bibinfo {pages} {392}
  (\bibinfo {year} {2004})}\BibitemShut {NoStop}%
\bibitem [{\citenamefont {Chapon}\ \emph {et~al.}(2006)\citenamefont {Chapon},
  \citenamefont {Radaelli}, \citenamefont {Blake}, \citenamefont {Park},\ and\
  \citenamefont {Cheong}}]{Chapon06}%
  \BibitemOpen
  \bibfield  {author} {\bibinfo {author} {\bibfnamefont {L.~C.}\ \bibnamefont
  {Chapon}}, \bibinfo {author} {\bibfnamefont {P.~G.}\ \bibnamefont
  {Radaelli}}, \bibinfo {author} {\bibfnamefont {G.~R.}\ \bibnamefont {Blake}},
  \bibinfo {author} {\bibfnamefont {S.}~\bibnamefont {Park}},\ and\ \bibinfo
  {author} {\bibfnamefont {S.-W.}\ \bibnamefont {Cheong}},\ }\href
  {https://doi.org/10.1103/PhysRevLett.96.097601} {\bibfield  {journal}
  {\bibinfo  {journal} {Phys. Rev. Lett.}\ }\textbf {\bibinfo {volume} {96}},\
  \bibinfo {pages} {097601} (\bibinfo {year} {2006})}\BibitemShut {NoStop}%
\bibitem [{\citenamefont {Giovannetti}\ \emph {et~al.}(2009)\citenamefont
  {Giovannetti}, \citenamefont {Kumar}, \citenamefont {Khomskii}, \citenamefont
  {Picozzi},\ and\ \citenamefont {van~den Brink}}]{Giovannetti09}%
  \BibitemOpen
  \bibfield  {author} {\bibinfo {author} {\bibfnamefont {G.}~\bibnamefont
  {Giovannetti}}, \bibinfo {author} {\bibfnamefont {S.}~\bibnamefont {Kumar}},
  \bibinfo {author} {\bibfnamefont {D.}~\bibnamefont {Khomskii}}, \bibinfo
  {author} {\bibfnamefont {S.}~\bibnamefont {Picozzi}},\ and\ \bibinfo {author}
  {\bibfnamefont {J.}~\bibnamefont {van~den Brink}},\ }\href
  {https://doi.org/10.1103/PhysRevLett.103.156401} {\bibfield  {journal}
  {\bibinfo  {journal} {Phys. Rev. Lett.}\ }\textbf {\bibinfo {volume} {103}},\
  \bibinfo {pages} {156401} (\bibinfo {year} {2009})}\BibitemShut {NoStop}%
\bibitem [{\citenamefont {Perez-Mato}\ \emph {et~al.}(2016)\citenamefont
  {Perez-Mato}, \citenamefont {Gallego}, \citenamefont {Elcoro}, \citenamefont
  {Tasci},\ and\ \citenamefont {Aroyo}}]{perez-mato16}%
  \BibitemOpen
  \bibfield  {author} {\bibinfo {author} {\bibfnamefont {J.~M.}\ \bibnamefont
  {Perez-Mato}}, \bibinfo {author} {\bibfnamefont {S.~V.}\ \bibnamefont
  {Gallego}}, \bibinfo {author} {\bibfnamefont {L.}~\bibnamefont {Elcoro}},
  \bibinfo {author} {\bibfnamefont {E.}~\bibnamefont {Tasci}},\ and\ \bibinfo
  {author} {\bibfnamefont {M.~I.}\ \bibnamefont {Aroyo}},\ }\href
  {https://doi.org/10.1088/0953-8984/28/28/286001} {\bibfield  {journal}
  {\bibinfo  {journal} {J. Phys.: Condens. Matter}\ }\textbf {\bibinfo {volume}
  {28}},\ \bibinfo {pages} {286001} (\bibinfo {year} {2016})}\BibitemShut
  {NoStop}%
\bibitem [{\citenamefont {Matsuda}\ \emph {et~al.}(2025)\citenamefont
  {Matsuda}, \citenamefont {Watanabe},\ and\ \citenamefont
  {Arita}}]{matsuda25}%
  \BibitemOpen
  \bibfield  {author} {\bibinfo {author} {\bibfnamefont {J.}~\bibnamefont
  {Matsuda}}, \bibinfo {author} {\bibfnamefont {H.}~\bibnamefont {Watanabe}},\
  and\ \bibinfo {author} {\bibfnamefont {R.}~\bibnamefont {Arita}},\ }\href
  {https://doi.org/10.1103/vgcs-bn8g} {\bibfield  {journal} {\bibinfo
  {journal} {Phys. Rev. Lett.}\ }\textbf {\bibinfo {volume} {134}},\ \bibinfo
  {pages} {226703} (\bibinfo {year} {2025})}\BibitemShut {NoStop}%
\bibitem [{\citenamefont {Anthony}\ \emph {et~al.}(2005)\citenamefont
  {Anthony}, \citenamefont {Bideaux}, \citenamefont {Bladh},\ and\
  \citenamefont {Nichols}}]{Anthony05}%
  \BibitemOpen
  \bibfield  {author} {\bibinfo {author} {\bibfnamefont {J.~W.}\ \bibnamefont
  {Anthony}}, \bibinfo {author} {\bibfnamefont {R.~A.}\ \bibnamefont
  {Bideaux}}, \bibinfo {author} {\bibfnamefont {K.~W.}\ \bibnamefont {Bladh}},\
  and\ \bibinfo {author} {\bibfnamefont {M.~C.}\ \bibnamefont {Nichols}},\
  }\href@noop {} {\emph {\bibinfo {title} {Handbook of Mineralogy}}}\ (\bibinfo
   {publisher} {Mineral Data Publishing},\ \bibinfo {year} {2005})\BibitemShut
  {NoStop}%
\bibitem [{\citenamefont {Corliss}\ \emph {et~al.}(1958)\citenamefont
  {Corliss}, \citenamefont {Elliott},\ and\ \citenamefont
  {Hastings}}]{Corliss1958}%
  \BibitemOpen
  \bibfield  {author} {\bibinfo {author} {\bibfnamefont {L.~M.}\ \bibnamefont
  {Corliss}}, \bibinfo {author} {\bibfnamefont {N.}~\bibnamefont {Elliott}},\
  and\ \bibinfo {author} {\bibfnamefont {J.~M.}\ \bibnamefont {Hastings}},\
  }\href {https://doi.org/10.1063/1.1723149} {\bibfield  {journal} {\bibinfo
  {journal} {J. Appl. Phys.}\ }\textbf {\bibinfo {volume} {29}},\ \bibinfo
  {pages} {391} (\bibinfo {year} {1958})}\BibitemShut {NoStop}%
\bibitem [{\citenamefont {Hastings}\ \emph {et~al.}(1959)\citenamefont
  {Hastings}, \citenamefont {Elliott},\ and\ \citenamefont
  {Corliss}}]{Hastings1959}%
  \BibitemOpen
  \bibfield  {author} {\bibinfo {author} {\bibfnamefont {J.~M.}\ \bibnamefont
  {Hastings}}, \bibinfo {author} {\bibfnamefont {N.}~\bibnamefont {Elliott}},\
  and\ \bibinfo {author} {\bibfnamefont {L.~M.}\ \bibnamefont {Corliss}},\
  }\href {https://doi.org/10.1103/physrev.115.13} {\bibfield  {journal}
  {\bibinfo  {journal} {Physical Review}\ }\textbf {\bibinfo {volume} {115}},\
  \bibinfo {pages} {13} (\bibinfo {year} {1959})}\BibitemShut {NoStop}%
\bibitem [{\citenamefont {Chattopadhyay}\ \emph {et~al.}(1991)\citenamefont
  {Chattopadhyay}, \citenamefont {Br\"uckel},\ and\ \citenamefont
  {Burlet}}]{Chattopadhyay1991PRB}%
  \BibitemOpen
  \bibfield  {author} {\bibinfo {author} {\bibfnamefont {T.}~\bibnamefont
  {Chattopadhyay}}, \bibinfo {author} {\bibfnamefont {T.}~\bibnamefont
  {Br\"uckel}},\ and\ \bibinfo {author} {\bibfnamefont {P.}~\bibnamefont
  {Burlet}},\ }\href {https://doi.org/10.1103/PhysRevB.44.7394} {\bibfield
  {journal} {\bibinfo  {journal} {Phys. Rev. B}\ }\textbf {\bibinfo {volume}
  {44}},\ \bibinfo {pages} {7394} (\bibinfo {year} {1991})}\BibitemShut
  {NoStop}%
\bibitem [{\citenamefont {Chatterji}\ \emph {et~al.}(2019)\citenamefont
  {Chatterji}, \citenamefont {Regnault}, \citenamefont {Ghosh},\ and\
  \citenamefont {Singh}}]{Chatterji2019}%
  \BibitemOpen
  \bibfield  {author} {\bibinfo {author} {\bibfnamefont {T.}~\bibnamefont
  {Chatterji}}, \bibinfo {author} {\bibfnamefont {L.~P.}\ \bibnamefont
  {Regnault}}, \bibinfo {author} {\bibfnamefont {S.}~\bibnamefont {Ghosh}},\
  and\ \bibinfo {author} {\bibfnamefont {A.}~\bibnamefont {Singh}},\ }\href
  {https://doi.org/10.1088/1361-648x/aafaeb} {\bibfield  {journal} {\bibinfo
  {journal} {J. Phys.: Condens. Matter}\ }\textbf {\bibinfo {volume} {31}},\
  \bibinfo {pages} {125802} (\bibinfo {year} {2019})}\BibitemShut {NoStop}%
\bibitem [{\citenamefont {Schick}\ \emph {et~al.}(2020)\citenamefont {Schick},
  \citenamefont {Ziman},\ and\ \citenamefont {Zhitomirsky}}]{Schick2020}%
  \BibitemOpen
  \bibfield  {author} {\bibinfo {author} {\bibfnamefont {R.}~\bibnamefont
  {Schick}}, \bibinfo {author} {\bibfnamefont {T.}~\bibnamefont {Ziman}},\ and\
  \bibinfo {author} {\bibfnamefont {M.~E.}\ \bibnamefont {Zhitomirsky}},\
  }\href {https://doi.org/10.1103/PhysRevB.102.220405} {\bibfield  {journal}
  {\bibinfo  {journal} {Phys. Rev. B}\ }\textbf {\bibinfo {volume} {102}},\
  \bibinfo {pages} {220405(R)} (\bibinfo {year} {2020})}\BibitemShut {NoStop}%
\bibitem [{sup()}]{suppl}%
  \BibitemOpen
  \href@noop {} {}\bibinfo {note} {See Supplementary Material for details of
  the uniaxial pressure dependence, field-cooled measurements of ferroelectric
  polarization, and a discussion of the effect of trapped charges.}\BibitemShut
  {Stop}%
\bibitem [{\citenamefont {Kimber}\ and\ \citenamefont
  {Chatterji}(2015)}]{Kimber2015}%
  \BibitemOpen
  \bibfield  {author} {\bibinfo {author} {\bibfnamefont {S.~A.~J.}\
  \bibnamefont {Kimber}}\ and\ \bibinfo {author} {\bibfnamefont
  {T.}~\bibnamefont {Chatterji}},\ }\href
  {https://doi.org/10.1088/0953-8984/27/22/226003} {\bibfield  {journal}
  {\bibinfo  {journal} {J. Phys.: Condens. Matter}\ }\textbf {\bibinfo {volume}
  {27}},\ \bibinfo {pages} {226003} (\bibinfo {year} {2015})}\BibitemShut
  {NoStop}%
\bibitem [{\citenamefont {Westrum}\ and\ \citenamefont
  {Gr{\o}nvold}(1970)}]{Westrum1970}%
  \BibitemOpen
  \bibfield  {author} {\bibinfo {author} {\bibfnamefont {E.~F.}\ \bibnamefont
  {Westrum}}\ and\ \bibinfo {author} {\bibfnamefont {F.}~\bibnamefont
  {Gr{\o}nvold}},\ }\href {https://doi.org/10.1063/1.1673563} {\bibfield
  {journal} {\bibinfo  {journal} {J. Chem. Phys.}\ }\textbf {\bibinfo {volume}
  {52}},\ \bibinfo {pages} {3820} (\bibinfo {year} {1970})}\BibitemShut
  {NoStop}%
\bibitem [{\citenamefont {Sidorov}\ \emph {et~al.}(2018)\citenamefont
  {Sidorov}, \citenamefont {Guo}, \citenamefont {Sun},\ and\ \citenamefont
  {Brazhkin}}]{Sidorov2018JETP}%
  \BibitemOpen
  \bibfield  {author} {\bibinfo {author} {\bibfnamefont {V.}~\bibnamefont
  {Sidorov}}, \bibinfo {author} {\bibfnamefont {J.}~\bibnamefont {Guo}},
  \bibinfo {author} {\bibfnamefont {L.}~\bibnamefont {Sun}},\ and\ \bibinfo
  {author} {\bibfnamefont {V.}~\bibnamefont {Brazhkin}},\ }\href
  {https://doi.org/10.1134/S0021364018050132} {\bibfield  {journal} {\bibinfo
  {journal} {JETP Letters}\ }\textbf {\bibinfo {volume} {107}},\ \bibinfo
  {pages} {311} (\bibinfo {year} {2018})}\BibitemShut {NoStop}%
\bibitem [{\citenamefont {Persson}\ \emph {et~al.}(2006)\citenamefont
  {Persson}, \citenamefont {Ceder},\ and\ \citenamefont
  {Morgan}}]{Persson2006}%
  \BibitemOpen
  \bibfield  {author} {\bibinfo {author} {\bibfnamefont {K.}~\bibnamefont
  {Persson}}, \bibinfo {author} {\bibfnamefont {G.}~\bibnamefont {Ceder}},\
  and\ \bibinfo {author} {\bibfnamefont {D.}~\bibnamefont {Morgan}},\ }\href
  {https://doi.org/10.1103/PhysRevB.73.115201} {\bibfield  {journal} {\bibinfo
  {journal} {Phys. Rev. B}\ }\textbf {\bibinfo {volume} {73}},\ \bibinfo
  {pages} {115201} (\bibinfo {year} {2006})}\BibitemShut {NoStop}%
\bibitem [{\citenamefont {Kimber}\ \emph {et~al.}(2014)\citenamefont {Kimber},
  \citenamefont {Salamat}, \citenamefont {Evans}, \citenamefont {Jeschke},
  \citenamefont {Muthukumar}, \citenamefont {Tomi{\'c}}, \citenamefont
  {Salvat-Pujol}, \citenamefont {Valent{\'i}}, \citenamefont {Kaisheva},
  \citenamefont {Zizak},\ and\ \citenamefont {Chatterji}}]{Kimber2014}%
  \BibitemOpen
  \bibfield  {author} {\bibinfo {author} {\bibfnamefont {S.~A.~J.}\
  \bibnamefont {Kimber}}, \bibinfo {author} {\bibfnamefont {A.}~\bibnamefont
  {Salamat}}, \bibinfo {author} {\bibfnamefont {S.~R.}\ \bibnamefont {Evans}},
  \bibinfo {author} {\bibfnamefont {H.~O.}\ \bibnamefont {Jeschke}}, \bibinfo
  {author} {\bibfnamefont {K.}~\bibnamefont {Muthukumar}}, \bibinfo {author}
  {\bibfnamefont {M.}~\bibnamefont {Tomi{\'c}}}, \bibinfo {author}
  {\bibfnamefont {F.}~\bibnamefont {Salvat-Pujol}}, \bibinfo {author}
  {\bibfnamefont {R.}~\bibnamefont {Valent{\'i}}}, \bibinfo {author}
  {\bibfnamefont {M.~V.}\ \bibnamefont {Kaisheva}}, \bibinfo {author}
  {\bibfnamefont {I.}~\bibnamefont {Zizak}},\ and\ \bibinfo {author}
  {\bibfnamefont {T.}~\bibnamefont {Chatterji}},\ }\href
  {https://doi.org/10.1073/pnas.1318543111} {\bibfield  {journal} {\bibinfo
  {journal} {Proc. Natl. Acad. Sci.}\ }\textbf {\bibinfo {volume} {111}},\
  \bibinfo {pages} {5106} (\bibinfo {year} {2014})}\BibitemShut {NoStop}%
\bibitem [{\citenamefont {Zhou}\ \emph {et~al.}(2024)\citenamefont {Zhou},
  \citenamefont {Zhong}, \citenamefont {Zhang}, \citenamefont {Kawaguchi},
  \citenamefont {Kadobayashi}, \citenamefont {Yu}, \citenamefont {Yue},\ and\
  \citenamefont {Hong}}]{Zhou2024}%
  \BibitemOpen
  \bibfield  {author} {\bibinfo {author} {\bibfnamefont {J.}~\bibnamefont
  {Zhou}}, \bibinfo {author} {\bibfnamefont {W.}~\bibnamefont {Zhong}},
  \bibinfo {author} {\bibfnamefont {H.}~\bibnamefont {Zhang}}, \bibinfo
  {author} {\bibfnamefont {S.}~\bibnamefont {Kawaguchi}}, \bibinfo {author}
  {\bibfnamefont {H.}~\bibnamefont {Kadobayashi}}, \bibinfo {author}
  {\bibfnamefont {X.}~\bibnamefont {Yu}}, \bibinfo {author} {\bibfnamefont
  {B.}~\bibnamefont {Yue}},\ and\ \bibinfo {author} {\bibfnamefont
  {F.}~\bibnamefont {Hong}},\ }\href
  {https://doi.org/10.1103/PhysRevB.110.165204} {\bibfield  {journal} {\bibinfo
   {journal} {Phys. Rev. B}\ }\textbf {\bibinfo {volume} {110}},\ \bibinfo
  {pages} {165204} (\bibinfo {year} {2024})}\BibitemShut {NoStop}%
\bibitem [{Note1()}]{Note1}%
  \BibitemOpen
  \bibinfo {note} {We note that even for $U_d$ = 7\protect \,eV (unusually
  large for a middle-transition-metal element like Mn), the experimental band
  gap is still underestimated by about 15\% (Fig.~\ref {fig:U-dependence}, b).
  While this discrepancy deserves a further investigation, it does not affect
  our central conclusions.}\BibitemShut {Stop}%
\bibitem [{\citenamefont {Giovannetti}\ and\ \citenamefont {van~den
  Brink}(2008)}]{Giovannetti08}%
  \BibitemOpen
  \bibfield  {author} {\bibinfo {author} {\bibfnamefont {G.}~\bibnamefont
  {Giovannetti}}\ and\ \bibinfo {author} {\bibfnamefont {J.}~\bibnamefont
  {van~den Brink}},\ }\href {https://doi.org/10.1103/PhysRevLett.100.227603}
  {\bibfield  {journal} {\bibinfo  {journal} {Phys. Rev. Lett.}\ }\textbf
  {\bibinfo {volume} {100}},\ \bibinfo {pages} {227603} (\bibinfo {year}
  {2008})}\BibitemShut {NoStop}%
\bibitem [{\citenamefont {Kresse}\ and\ \citenamefont
  {Furthm{\"{u}}ller}(1996{\natexlab{a}})}]{vasp_1}%
  \BibitemOpen
  \bibfield  {author} {\bibinfo {author} {\bibfnamefont {G.}~\bibnamefont
  {Kresse}}\ and\ \bibinfo {author} {\bibfnamefont {J.}~\bibnamefont
  {Furthm{\"{u}}ller}},\ }\href {https://doi.org/10.1103/PhysRevB.54.11169}
  {\bibfield  {journal} {\bibinfo  {journal} {Phys. Rev. B}\ }\textbf {\bibinfo
  {volume} {54}},\ \bibinfo {pages} {11169} (\bibinfo {year}
  {1996}{\natexlab{a}})}\BibitemShut {NoStop}%
\bibitem [{\citenamefont {Kresse}\ and\ \citenamefont
  {Furthm{\"{u}}ller}(1996{\natexlab{b}})}]{vasp_2}%
  \BibitemOpen
  \bibfield  {author} {\bibinfo {author} {\bibfnamefont {G.}~\bibnamefont
  {Kresse}}\ and\ \bibinfo {author} {\bibfnamefont {J.}~\bibnamefont
  {Furthm{\"{u}}ller}},\ }\href {https://doi.org/10.1016/0927-0256(96)00008-0}
  {\bibfield  {journal} {\bibinfo  {journal} {Comput. Mater. Sci.}\ }\textbf
  {\bibinfo {volume} {6}},\ \bibinfo {pages} {15} (\bibinfo {year}
  {1996}{\natexlab{b}})}\BibitemShut {NoStop}%
\bibitem [{\citenamefont {Kresse}\ and\ \citenamefont
  {Joubert}(1999)}]{vasp_paw}%
  \BibitemOpen
  \bibfield  {author} {\bibinfo {author} {\bibfnamefont {G.}~\bibnamefont
  {Kresse}}\ and\ \bibinfo {author} {\bibfnamefont {D.}~\bibnamefont
  {Joubert}},\ }\href {https://doi.org/10.1103/PhysRevB.59.1758} {\bibfield
  {journal} {\bibinfo  {journal} {Phys. Rev. B}\ }\textbf {\bibinfo {volume}
  {59}},\ \bibinfo {pages} {1758} (\bibinfo {year} {1999})}\BibitemShut
  {NoStop}%
\bibitem [{\citenamefont {Perdew}\ \emph {et~al.}(2008)\citenamefont {Perdew},
  \citenamefont {Ruzsinszky}, \citenamefont {Csonka}, \citenamefont {Vydrov},
  \citenamefont {Scuseria}, \citenamefont {Constantin}, \citenamefont {Zhou},\
  and\ \citenamefont {Burke}}]{perdew2008restoring}%
  \BibitemOpen
  \bibfield  {author} {\bibinfo {author} {\bibfnamefont {J.~P.}\ \bibnamefont
  {Perdew}}, \bibinfo {author} {\bibfnamefont {A.}~\bibnamefont {Ruzsinszky}},
  \bibinfo {author} {\bibfnamefont {G.~I.}\ \bibnamefont {Csonka}}, \bibinfo
  {author} {\bibfnamefont {O.~A.}\ \bibnamefont {Vydrov}}, \bibinfo {author}
  {\bibfnamefont {G.~E.}\ \bibnamefont {Scuseria}}, \bibinfo {author}
  {\bibfnamefont {L.~A.}\ \bibnamefont {Constantin}}, \bibinfo {author}
  {\bibfnamefont {X.}~\bibnamefont {Zhou}},\ and\ \bibinfo {author}
  {\bibfnamefont {K.}~\bibnamefont {Burke}},\ }\href
  {https://doi.org/10.1103/PhysRevLett.100.136406} {\bibfield  {journal}
  {\bibinfo  {journal} {Phys. Rev. Lett.}\ }\textbf {\bibinfo {volume} {100}},\
  \bibinfo {pages} {136406} (\bibinfo {year} {2008})}\BibitemShut {NoStop}%
\bibitem [{\citenamefont {Dudarev}\ \emph {et~al.}(1998)\citenamefont
  {Dudarev}, \citenamefont {Botton}, \citenamefont {Savrasov}, \citenamefont
  {Humphreys},\ and\ \citenamefont {Sutton}}]{dudarev1998electron}%
  \BibitemOpen
  \bibfield  {author} {\bibinfo {author} {\bibfnamefont {S.~L.}\ \bibnamefont
  {Dudarev}}, \bibinfo {author} {\bibfnamefont {G.~A.}\ \bibnamefont {Botton}},
  \bibinfo {author} {\bibfnamefont {S.~Y.}\ \bibnamefont {Savrasov}}, \bibinfo
  {author} {\bibfnamefont {C.}~\bibnamefont {Humphreys}},\ and\ \bibinfo
  {author} {\bibfnamefont {A.~P.}\ \bibnamefont {Sutton}},\ }\href
  {https://doi.org/10.1103/PhysRevB.57.1505} {\bibfield  {journal} {\bibinfo
  {journal} {Phys. Rev. B}\ }\textbf {\bibinfo {volume} {57}},\ \bibinfo
  {pages} {1505} (\bibinfo {year} {1998})}\BibitemShut {NoStop}%
\bibitem [{\citenamefont {Togo}\ \emph {et~al.}(2023)\citenamefont {Togo},
  \citenamefont {Chaput}, \citenamefont {Tadano},\ and\ \citenamefont
  {Tanaka}}]{phonopy_1}%
  \BibitemOpen
  \bibfield  {author} {\bibinfo {author} {\bibfnamefont {A.}~\bibnamefont
  {Togo}}, \bibinfo {author} {\bibfnamefont {L.}~\bibnamefont {Chaput}},
  \bibinfo {author} {\bibfnamefont {T.}~\bibnamefont {Tadano}},\ and\ \bibinfo
  {author} {\bibfnamefont {I.}~\bibnamefont {Tanaka}},\ }\href
  {https://doi.org/10.1088/1361-648X/acd831} {\bibfield  {journal} {\bibinfo
  {journal} {J. Phys.: Condens. Matter}\ }\textbf {\bibinfo {volume} {35}},\
  \bibinfo {pages} {353001} (\bibinfo {year} {2023})}\BibitemShut {NoStop}%
\bibitem [{\citenamefont {Togo}(2023)}]{phonopy_2}%
  \BibitemOpen
  \bibfield  {author} {\bibinfo {author} {\bibfnamefont {A.}~\bibnamefont
  {Togo}},\ }\href {https://doi.org/10.7566/JPSJ.92.012001} {\bibfield
  {journal} {\bibinfo  {journal} {J. Phys. Soc. Jpn.}\ }\textbf {\bibinfo
  {volume} {92}},\ \bibinfo {pages} {012001} (\bibinfo {year}
  {2023})}\BibitemShut {NoStop}%
\bibitem [{\citenamefont {Stokes}\ and\ \citenamefont
  {Hatch}(2005)}]{stokes2005findsym}%
  \BibitemOpen
  \bibfield  {author} {\bibinfo {author} {\bibfnamefont {H.~T.}\ \bibnamefont
  {Stokes}}\ and\ \bibinfo {author} {\bibfnamefont {D.~M.}\ \bibnamefont
  {Hatch}},\ }\href {https://doi.org/10.1107/S0021889804031528} {\bibfield
  {journal} {\bibinfo  {journal} {J. Appl. Crystallogr.}\ }\textbf {\bibinfo
  {volume} {38}},\ \bibinfo {pages} {237} (\bibinfo {year} {2005})}\BibitemShut
  {NoStop}%
\bibitem [{\citenamefont {Resta}(1994)}]{resta1994macroscopic}%
  \BibitemOpen
  \bibfield  {author} {\bibinfo {author} {\bibfnamefont {R.}~\bibnamefont
  {Resta}},\ }\href {https://doi.org/https://doi.org/10.1103/RevModPhys.66.899}
  {\bibfield  {journal} {\bibinfo  {journal} {Rev. Mod. Phys.}\ }\textbf
  {\bibinfo {volume} {66}},\ \bibinfo {pages} {899} (\bibinfo {year}
  {1994})}\BibitemShut {NoStop}%
\end{thebibliography}
\end{document}